\documentclass[twocolumn,pra,showpacs,superscriptaddress]{revtex4-1}
\usepackage{hyperref}
\usepackage{graphicx}% Include figure files
\usepackage{dcolumn}% Align table columns on decimal point
\usepackage{bm}
\usepackage{tikz}
\usepackage{amsmath}
\usetikzlibrary{decorations.markings}
\usetikzlibrary{decorations.pathmorphing}

\DeclareMathOperator{\Tr}{Tr}

\begin{document}

% \preprint{APS/123-QED}

\title{Entanglement witness game}

\author{Xing Chen}

\affiliation{Institute of Physics, Beijing National Laboratory for
  Condensed Matter Physics, Chinese Academy of Sciences, Beijing
  100190, China}

\affiliation{School of Physical Sciences, University of Chinese
  Academy of Sciences, Beijing 100190, China}

\author{Xueyuan Hu}

\affiliation{School of Information Science and
  Engineering, and Shandong Provincial Key Laboratory of Laser
  Technology and Application, Shandong University, Jinan 250100,
  China}

\author{D.~L. Zhou}

\email{zhoudl72@iphy.ac.cn}

\affiliation{Institute of Physics, Beijing National Laboratory for
  Condensed Matter Physics, Chinese Academy of Sciences, Beijing
  100190, China}

\affiliation{School of Physical Sciences, University of Chinese
  Academy of Sciences, Beijing 100190, China}

\begin{abstract}
  Motivated by Buscemi's semi-quantum nonlocal game [PRL 108, 200401
  (2012)], We propose an entanglement witness game, a quantum game
  based on entanglement witness. Similar as the
  semi-quantum nonlocal game, the existence of entanglement shared by
  the players is necessary and sufficient for obtaining a positive
  average payoff in our entanglement witness game. Two explicit examples are
  constructed to demonstrate how to play our entanglement witness game and its
  relations with the CHSH nonlocal game.
\end{abstract}


\maketitle

%\tableofcontents

\section{Introduction}

Entanglement is a useful resource in quantum information processes.
For example, quantum teleportation is impossible without shared remote
entanglement. What's the property of entanglement makes it
indispensable in these tasks? It is believed that entanglement has
some intriguing nonlocality, which can be revealed by violating Bell
inequalities based on local Hidden variables. In 1989, however, R.F.
Werner \cite{wer} discovered that there are some entanglement states
satisfy all Bell inequalities, in other words, the
correlations\cite{bel}\cite{cla} in these entangled states can be
explained by the Hidden variable model.

Since Werner's discovery, a lot of works aimed to explore the subtle
differences between the nonlocality in entanglement and that revealed
by violating Bell inequalities \cite{hor}\cite{mas}\cite{met}. For
instance, the following problem is raised \cite{pop}\cite{hor3}: are
all the entangled states useful in quantum teleportation? This leads
to the discovery of the concept of bound entanglement and the
understanding of the role of entanglement played in quantum
teleportation.

In another direction, the relation between nonlocal games and Bell
inequalities is built: For any Bell inequality, we can build a
nonlocal game \cite{sil}\cite{cle} which makes the states violating
Bell inequality have a positive average payoff. According to Werner's
discovery, the entanglement states satisfying all Bell inequalities
will not have a positive average payoff in any such type of nonlocal
game. This makes F. Buscemi put forward a revised nonlocal game
\cite{bus}, called semi-quantum nonlocal game, in which the questions
sent to the players are allowed to be represented by nonorthogonal
quantum states. In the semi-quantum nonlocal game, any entangled state
can obtain a positive average payoff while all the separable states
can not.

In 2013, two groups \cite{bra}\cite{cav} realize that the semi-quantum
nonlocal game can be used to be measurement-device-independent
entanglement witness\cite{hor}. In particular, it shows how to use any
entanglement witness to design a semi-quantum nonlocal
game~\cite{bra}.

Motivated by the semi-quantum nonlocal game, we propose a simpler
quantum game based on entanglement witness, in which only classical
communications between the referee and the players are allowed, and the target of our game is the
same as\cite{bus}: any entangled state can obtain higher payoff than
all separable states.

Our paper is organized as follows. In section II we briefly review
nonlocal game\cite{cle} and semi-quantum nonlocal game \cite{bus}. In
section III, the main part of this paper, we propose our entanglement witness game,
for simplicity, we only give the bipartite and tripartite
scenario of our game, but our game can be easily extend to
multipartite situations. In section IV, we give explicit two-qubit
examples to show how to play our entanglement witness game, and it also reveals the
relation of our entanglement witness game with the previous one. Finally, a
summary is given.

\section{Nonlocal game and Semi-quantum nonlocal game}

\begin{figure}[hbt!]
\centering
\includegraphics[width=6.6cm]{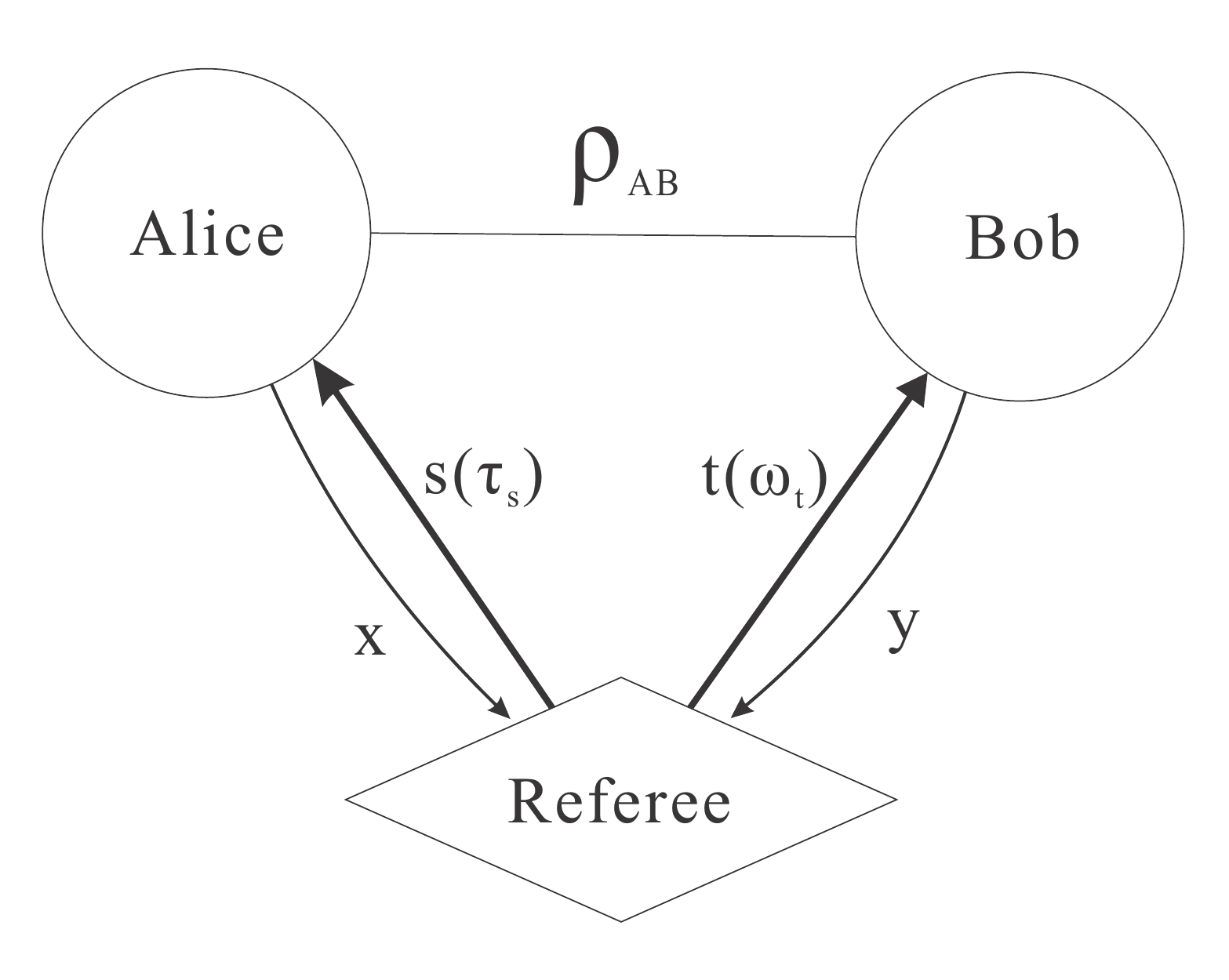}
\caption{Nonlocal (Semi-quantum nonlocal) game}\label{fig1}
\end{figure}

Before discussing our entanglement witness game, we review the procedures of
nonlocal game \cite{cle} and semi-quantum nonlocal game, which
are % The two games are very similar, as
shown in Fig.~\ref{fig1}.

In these two games, there are two players, Alice and Bob, and one
referee. The referee sends two questions, which are represented by two
quantum states $\tau_{s}$ and $\omega_{t}$, to Alice and Bob
respectively with a probability distribution $\Pi(s,t)$. After
obtaining the questions, Alice and Bob make some local operations to
give their answers $x$ and $y$ respectively with a conditional
probability distribution $V(x,y|s,t)$. Then the referee will pay them
according to a given payoff function $\wp(x,y,s,t)$.

In the games, Alice and Bob share a bipartite quantum state
$\rho_{AB}$, which can be regarded as the resource of nonlocality. In
other words, Alice and Bob make use of the nonlocality of the state
$\rho_{AB}$ to get the payoff as much as possible.

More precisely, Alice and Bob try to choose an optimal $V(x,y|s,t)$ to
maximize the average payoff:
\begin{equation*}
  \mathbf{P}(\rho_{AB},G_{n}) = \max \sum_{s,t,x,y} \Pi(s,t) V(x,y|s,t)
  \wp(x,y,s,t).
\end{equation*}
There are two basic differences between nonlocal game and
semi-nonlocal game. First, $\{\tau_{s}\}$ and $\{\omega_{t}\}$ are
distinguishable for Alice and Bob, i.e., they are orthogonal in the
nonlocal game; however, we require the completeness of the states in
the operator space but not the orthogonality in the state space in the
semi-nonlocal game.

Second, the criteria to win the games are different. In nonlocal
game, Alice and Bob win if the payoff $ \textbf{P}(\rho_{AB},G_{n})$
outweighs any LHV (Local Hidden Variables) states and lose otherwise.
In the semi-nonlocal game, Alice and Bob win if the payoff is larger
than that from any separable bipartite quantum state.

Here it is worthy to emphasize that, before the start of the games,
Alice and Bob are allowed to communicate with each other to work out
some strategies, then the communication is not permitted during the
processes of the games.

In the two games, without using deterministic strategies, the
conditional probabilities $V(x,y|s,t)$ are given based on suitable
quantum measurements.

In nonlocal game, $V$ is
\begin{equation}
  V(x,y|s,t) = \Tr(\rho_{AB} A^x_s\otimes B^y_t),
\end{equation}
where $A^x_s$ and $B^y_t$ are the measurement make by Alice and Bob
according to their received labels $s$ and $t$ of $\{\tau^{s}\}$ and
$\{\omega^{t}\}$. Since the construction of nonlocal game is based on
Bell inequality~\cite{cle}, from the form of Bell inequality(like CHSH
inequality ) we can easily get the desired $A^x_s$ and $B^y_t$.

In semi-quantum nonlocal game, for any entangled state, the
distribution $V$ is written as follows
\begin{equation}
 V(x,y|s,t) = \Tr(P^x_{AA_0}\otimes P^y_{BB_0}
 \tau^s_{A_0}\otimes\rho_{AB}\otimes\omega^t_{B_0}),
\end{equation}
where $P^x_{AA_0}$ is the POVM Alice makes on the entangled state
$\rho_{AB}$ and her received state $\tau^s_{A_0}$, and $P^y_{BB_0}$ is
Bob's POVM. $V$ of entangled states is out of reach for any separable
states, which makes the payoff of any entangled state larger than
separable states.

\section{Our entanglement witness game}

In this section, we propose a quantum game to use entanglement as a
resource. To simplify our notations, we assume that $\rho_{AB}$ is a
two-qubit state.

The procedure of our entanglement witness game is as follows.

First, the referee randomly sends the labels $s,t\in\{0,1,2,3\}$ to
Alice and Bob with probability $\Pi(s,t)$;

Second, according to labels $s,t$ received from the referee,
Alice and Bob do the following measurements
\begin{equation}
    0\rightarrow \textbf{I}_{2\times2}\quad 1\rightarrow \sigma_x
    \quad 2\rightarrow \sigma_y \quad 3\rightarrow \sigma_z
    \label{measurement}
\end{equation}
and then Alice and Bob return the measurement results
$a,b\in\{-1,+1\}$ back
to the referee;

Third, the referee pays them according to the payoff function
$\wp(a,b,s,t)=-w_{s,t}~ab/\Pi(s,t)$, where the
referee chooses $\{w_{s,t}\}$ such that
\begin{equation}
  W=\sum_{s,t=0}^3 w_{s,t}\sigma_{s}\otimes\sigma_{t}
  \label{witness}
\end{equation}
is an entanglement witness of $\rho_{AB}$ in the case when $\rho_{AB}$
is entangled, where $\sigma_0=\textbf{I}_{2\times2}$, $\sigma_{1,2,3}$ are Pauli matrices.

Since $W$ is the entanglement witness of the entangled state
$\rho_{AB}$, we have \cite{hor}\cite{guh}
\begin{align}
  \Tr(\rho_{AB} W) & < 0, \\
  \Tr(\sigma_{AB} W) & \ge 0,
\end{align}
where $\sigma_{AB}$ is any separable state. Notice that the average
payoff
\begin{align}
  \label{eq:1}
   \textbf{P} & = \sum_{s,t,a,b} \Pi(s,t) V(a,b|s,t)
               \wp(a,b,s,t) \nonumber\\
  & = - \Tr(\rho_{AB} W),
\end{align}
which ensures that only when $\rho_{AB}$ is entangled the average
payoff is positive. Further more, because there always exists an
entanglement witness for any entangled state $\rho_{AB}$, the payoff
function based on the entanglement witness can be constructed, which
makes any entanglement can be distinguished in our game.

For the referee, the process of our entanglement witness game is a stochastic
tomography of a two-qubit state. When an ensemble of a two-qubit state
is input, the two-qubit state can be reconstructed by the referee with
the answers from Alice and Bob. The entanglement witness is involved
in our game through a proper choice of the payoff function. Because
the entanglement witness is relative to a given entangled state, the
payoff functions may be different for different entangled states
shared by the players. In particular, for any entanglement state, we
only need to adopt a proper entanglement witness to design the payoff
function for our entanglement witness game.

% Since only classical questions are sent to the two players, our local
% game belongs to nonlocal game. There are two obvious differences
% between the traditional nonlocal game and our nonlocal game: First,
% there are $2$ unfixed measurement operators for each player, while in
% our nonlocal game, there are $4$ fixed measurement operators for each
% player. Second, the criteria to win in our nonlocal game is the same
% as the semi-nonlocal game, but not as that adopt for the traditional
% nonlocal game.

Different from the traditional nonlocal game, where the player can win
only when they share Bell nonlocal state, our entanglement witness game gives a
positive average payoff if the players share a proper entangled state
(which may not violate any Bell inequality). The reason for this is
that, in the traditional game, the player can choose any two local
measurements, while in our game, they are restricted to the four fixed
measurement operators.

Our entanglement witness game can be easily extend to multipartite situations. We
use three-qubit scenario to illustrate this. For tripartite games,
there are three players, Alice, Bob and Charlie. The entanglement
witness of three-qubit entangled states $\rho_{ABC}$ can be decomposed
as
\begin{equation}
  W=\sum_{i,j,k=0}^3w_{i,j,k}\sigma_i\otimes\sigma_j\otimes\sigma_k
\end{equation}
The same as bipartite scenario, the referee send Alice, Bob and
Charlie four labels $i,j,k\in\{0,1,2,3\}$, and the players do the
measurement of $\sigma_{i}\otimes\sigma_{j}\otimes\sigma_{k}$. Then
they send back their binary-outcome results
$a,b,c\in\{-1,+1\}$, the referee pay them according to the
payoff function $-w_{i,j,k}~abc/\Pi(i,j,k)$. Then the average payoff
would be
\begin{equation}
  \textbf{P}=-\Tr(\rho_{ABC}W)
\end{equation}
Only when $\rho_{ABC}$ is a tripartite entangled state, the payoff
is larger than zero.

\section{Explicit Examples: Two-qubit case}

In this section, we give two explicit examples to show how to play our entanglement witness game,
and show that Alice and Bob are required to be honest in our game.

In the first example, Alice and Bob share the Werner state~\cite{wer}:
\begin{equation}
\label{werner}
  \rho_z = \frac{1-z}{4} \textbf{I} + z|\psi^+\rangle\langle\psi^+|,
\end{equation}
where $I=\sigma_{0}\otimes\sigma_{0}$,
$|\psi^+\rangle=(|00\rangle+|11\rangle)/\sqrt{2}$, $z\in[0,1]$. The
entanglement witness of this Werner state is\cite{hyl}
\begin{equation}
\label{werner-witness}
W =\frac{1}{\sqrt{3}} (\textbf{I} -
  \sigma_1\otimes\sigma_1 + \sigma_2\otimes\sigma_2 -
  \sigma_3\otimes\sigma_3).
\end{equation}
% The referee send Alice and Bob four labels, and then Alice and Bob
% measure with corresponding operators
% $\textbf{I}, \sigma_1, \sigma_2,\sigma_3$. For this Werner state, only
% when Alice and Bob receive the same labels $00,11,22,33$, the
% correlation between their measurement results are not zero.
The factor $\frac{1}{ \sqrt{3}}$ in $W$ makes the projection of $W$
into the linear space expanded by
$\{\sigma_{i}\otimes\sigma_{i}, i\in\{1,2,3\}\}$ is a normalized
vector if we define the inner product $\langle A,B \rangle=\Tr(AB)$
for any two Hermitian operators.

In fact, any two-qubit density matrix $\rho$ is an element in the
linear space expanded by
$\{\sigma_{i}\otimes\sigma_{j}, i,j\in{0,1,2,3}\}$. Since
$\Tr(\rho)=1$, the component of $I$ is $\frac{1}{4}$. Thus all the
two-qubit density matrices form a convex body in $15$ dimensional
linear space, where a general entanglement witness for two-qubit
states lies in according to convex analysis~\cite{rock}. Usually, it
is more convenient to study its subspaces, e.g., a $3$ dimensional
subspace, or even a $2$ dimensional subspace.

\begin{figure}[htbp]
  \centering
  \includegraphics[width=\linewidth]{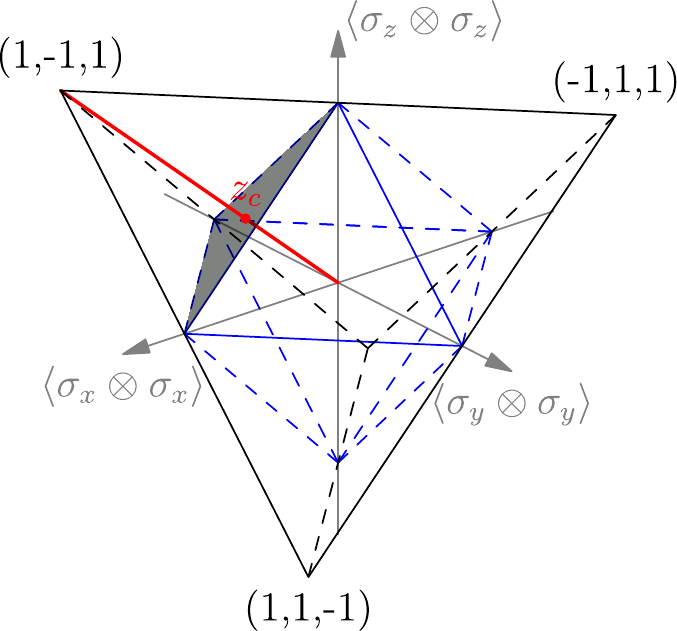}
  \caption{Entanglement witness $W$ and the projections of two-qubit
    states. The regular tetrahedron formed by four black lines is the
    numerical range of two-qubit states,  and the regular octahedron
    formed by twelve blue lines is the numerical range for two-qubit
    separable states. $\Tr(\rho W)=0$ corresponds to a gray
    hyperplane, which is also a surface of the regular octahedron. The
    projection of the Werner state $\rho_{z}$ is denoted by a
    red line, which intersects with the hyperplane at $z_{c}=\frac{1}{3}$.  }
  \label{fig2}
\end{figure}

Here we focus on the $3$ dimensional subspace expanded by
$\{\sigma_{i}\otimes\sigma_{i}, i\in\{1,2,3\}\}$, and the projections
of two-qubit states and the entanglement witness are shown in
Fig.~\ref{fig2}. The coordinate for any two-qubit state $\rho$ in this
subspace is given by
$(\langle \sigma_{x} \otimes \sigma_{x} \rangle, \langle \sigma_{y}
\otimes \sigma_{y} \rangle, \langle \sigma_{z} \otimes \sigma_{z}
\rangle )$ with $\langle \cdot \rangle = \Tr(\rho\; \cdot)$. In this subspace,
the numerical range~\cite{chen2016} for all two-qubit states is the
regular tetrahedron with vertexes
$\{(1,-1,1),(-1,1,1),(1,1,-1),(-1,-1,-1)\}$, and the numerical range
for all separable two-qubit states is the octahedron with vertexes
$\{(\pm 1,0,0),(0, \pm 1, 0), (0,0,\pm 1)\}$. The gray hyperplane is
defined by $\Tr(\rho W)=0$. The Werner state $\rho_{z}$ is
represented by a red line, which intersects with the hyperplane at
$z_{c}=\frac{1}{3}$.

According to the rule of our entanglement witness game, the average payoff
for the Werner state is
\begin{equation}
 \textbf{P}=-\sum_{i,j}w_{i,j} \Tr(\rho_{z} \sigma_i\otimes\sigma_j)
 =\frac{3z-1}{ \sqrt{3}},
\end{equation}
which is  the geometric distance between the state and
the hyperplane. As we know, $\rho_z$ is entangled only when $z>z_{c}=1/3$.
The above equation shows that when $z>z_{c}$ the payoff is larger than
zero, which shows that the entanglement in the Werner state is a
useful resource for our entanglement witness game.

Here let us explain why Alice and Bob are required to be honest in our game by an explicit construction. Assume that Alice and Bob share
the classical correlated state
\begin{equation}
  \label{eq:10}
  \rho_{A1,A2,A3,B1,B2,B3}=\rho_{A1,B1}\otimes\rho_{A2,B2}\otimes\rho_{A3,B3},
\end{equation}
with
\begin{align}
  \rho_{A1,B1} & = \frac{I^{A1,B1} + \sigma_{x}^{A1}\otimes\sigma_{x}^{B1}}{4}, \\
  \rho_{A2,B2} & = \frac{I^{{A2,B2}} - \sigma_{y}^{A2}\otimes\sigma_{y}^{B2}}{4}, \\
  \rho_{A3,B3} & = \frac{I^{A3,B3} + \sigma_{z}^{A3}\otimes\sigma_{z}^{B3}}{4}.
\end{align}
Then they can simulate the correlations in the entangled state
$|\psi^{+}\rangle$ in a scheme discussed by them before the game
starts as follows. When the referee sends $1$, $2$, and $3$ to Alice
(Bob), Alice (Bob) measures
$\sigma_{x}^{A1}, \sigma_{y}^{A2}, \sigma_{z}^{A3}$
($\sigma_{x}^{B1}, \sigma_{y}^{B2}, \sigma_{z}^{B3}$)
respectively. Alice and Bob return the measurement results $\pm 1$ to
the referee. Then Alice and Bob will get the maximal average payoff
$2\sqrt{3}/3$ in our game based on the entanglement witness
$W$. This construction shows that the players can get a positive
average payoff without shared entanglement if they are allowed to be
dishonest in our entanglement witness game.

Another example is from the CHSH inequality\cite{cla}, in this example
we will show the relation between the CHSH inequality and our entanglement
witness game. This relation is based on the relation between CHSH inequality and
entanglement witness \cite{hyl}. The CHSH inequality
\begin{equation}
  \label{eq:2}
 -2\leq\langle A\otimes B + A^{\prime} \otimes
  B + A\otimes B^{\prime} - A^{\prime} \otimes B^{\prime}\rangle\le 2
\end{equation}
gives entanglement witness
\begin{equation}
  \label{eq:3}
  W^{\pm}_{\text{CHSH}} = 2\textbf{I} \pm\-  (A\otimes B + A^{\prime} \otimes
  B + A\otimes B^{\prime} - A^{\prime} \otimes B^{\prime}).
\end{equation}

For the Werner state $\rho_{z}$, especially the state when $z=1$,
Alice and Bob randomly choose their spin observables:
\begin{equation}
\begin{split}
  & A=\sigma_x \quad or \quad A'=\sigma_z, \\
  & B =-\frac{\sigma_x+\sigma_z}{\sqrt{2}}\quad or \quad
  B'=\frac{\sigma_z-\sigma_x}{\sqrt{2}}.
\end{split}\label{eq:9}
\end{equation}
Now the CHSH entanglement witnesses become
\begin{equation}
  \label{eq:4}
  W_{\text{CHSH}} = \textbf{I} - \frac{\sigma_{x} \otimes \sigma_{x} +
    \sigma_{z} \otimes \sigma_{z} }{\sqrt{2}},
\end{equation}
where a normalized factor $1/2$ is contained.

\begin{figure}[htbp]
  \centering
  \includegraphics[width=\linewidth]{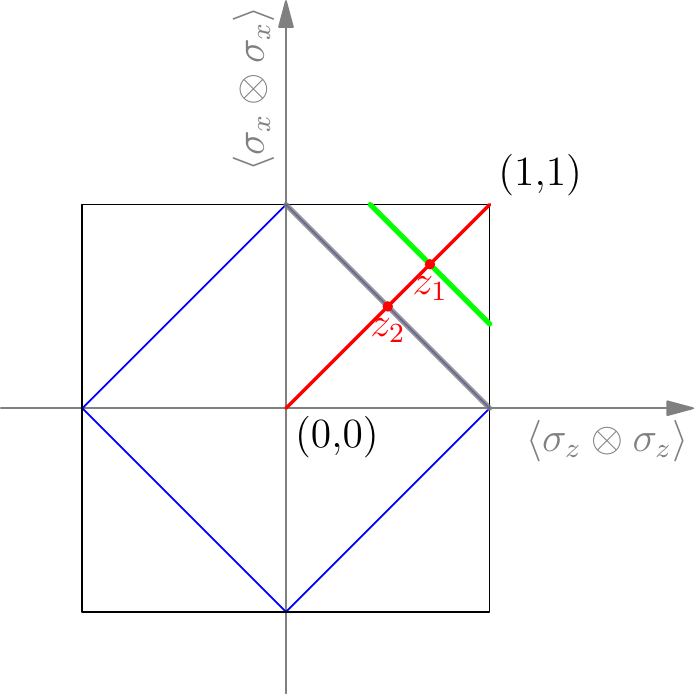}
  \caption{Entanglement witnesses $W_{\text{CHSH}}$, $\bar{W}$ and the
    projections of two-qubit states. The black square is the numerical
    range of two-qubit states, and the blue square is the numerical
    range for two-qubit separable states.
    $\Tr(\rho W_{\text{CHSH}})=0$ corresponds to the green line, and
    $\Tr(\rho \bar{W})=0$ corresponds to the gray line. The projection
    of the Werner state $\rho_{z}$ is denoted by a red line, which
    intersects with the above two lines at $z_{1}=\frac{\sqrt{2}}{2}$
    and $z_{2}=1/2$. }
  \label{fig3}
\end{figure}

Because $W_{\text{CHSH}}$ is related with
$\{\sigma_{x}\otimes\sigma_{x}, \sigma_{z}\otimes\sigma_{z}\}$, which
motivates us to consider the $2$ dimensional subspace expanded by
them. The similar numerical results are shown in Fig.~\ref{fig3}.
Now the numerical range for all two-qubit states is the black square,
and the numerical range for all separable two-qubit states is the blue
square. The hyperplanes $\Tr(\rho W_{\text{CHSH}})=0$ is the green
line, and the Werner state $\rho_{z}$ is denoted as the red line,
which intersects with the green line at $z_{1}=\frac{\sqrt{2}}{2}$.

In our entanglement witness game, the average payoff is
\begin{equation}
  \label{eq:6}
  \textbf{P} = - \Tr(\rho_{z} W_{CHSH}) = \sqrt{2} z - 1,
\end{equation}
which is the geometric distance from a Werner state to the green line.
When $P>0$, we require that $z>\sqrt{2}/2$. As expected,
$P(\rho_{z})>0$ implies $\rho_{z}$ is entangled. However, when
$1/3<z\leq\sqrt{2}/2$, $\rho_{z}$ is entangled but not detected in our entanglement witness game based on the CHSH inequality, which is consistent with \cite{hor2}\cite{wis}.

In fact, in our entanglement witness game, we strengthen the CHSH entanglement witness as
\begin{equation}
  \label{eq:7}
  \bar{W} = \frac{1}{\sqrt{2}} (\textbf{I} -
  \sigma_1\otimes\sigma_1 - \sigma_3\otimes\sigma_3),
\end{equation}
which corresponds to a gray line in Fig.~\ref{fig3} defined by
$\Tr(\rho \bar{W})$. The Werner state interact with the gray line at
$z_{2}=\frac{1}{2}$.

Then for the Werner state, the average payoff
\begin{equation}
  \label{eq:8}
  \textbf{P} = - \Tr(\rho_{z} \bar{W}) = \frac{\sqrt{2}}{2} (2z - 1).
\end{equation}
For  $A,A^{\prime},B,B^{\prime}$ given by Eq.~\eqref{eq:9}, the Bell inequality as
entanglement witness can be strengthened as
\begin{equation}
  -\sqrt{2}\leq\langle A\otimes B + A^{\prime} \otimes
  B + A\otimes B^{\prime} - A^{\prime} \otimes B^{\prime}\rangle\le
  \sqrt{2}.
\end{equation}
Even after the strengthen, only the Werner state for $z>\frac{1}{2}$
is entanglement witnessed, and the entangled Werner state for
$\frac{1}{3}<z\le \frac{1}{2}$ is not detected in our entanglement witness game.

From the above two typical examples of our entanglement witness game, we find that
the higher dimension of the consider subspace, more entangled states
are useful for our entanglement witness game. An interesting question to be
explored in future arises: what is the lowest dimension of the
subspace to make all entangled states are useful in our entanglement witness game?

\section{Conclusion}

In this paper, we proposed an entanglement witness game, which ensures any
entangled state can have a positive payoff while separable states can
not. The process of our entanglement witness game, in the viewpoint of the
referee, is a stochastic local quantum state tomography, which makes
the referee has the capacity to obtain the entangled state shared by
the players asymptotically. It is worthy to point out if the players
are dishonest, they may get positive average payoff without shared
entanglement, which is not robust as in the semi-nonlocal game.

Compared with the nonlocal game from the Bell inequalities, any
entangled state is useful in our entanglement witness game, while in the
traditional one, only the entangled states violating Bell inequalities
has advantage over separable states. Compared with the semi-quantum
nonlocal game, we do not need quantum channels to transfer
non-orthogonal quantum states from the referee to the players. In
addition, the procedure is relative simpler than the previous ones,
and it is more directly based on entanglement witness. Our entanglement witness game
 can use any entangled state as a resource by adopting its
entanglement witness to design the payoff function, and it can
be implemented similarly as the entanglement witness
experiments~\cite{barb}\cite{bour}.

We hope our work will increase our understandings on Bell nonlocality
and entanglement through different quantum games, and promote their
applications in quantum information processes.

\begin{acknowledgments}
 This work is supported by NSF of China (Grant No. 11475254)(BNKBRSF
 of China (Grant No. 2014CB921202), and The National Key Research and
 Development Program of China(Grant No. 2016YFA0300603).
\end{acknowledgments}

\bibliography{newnonlocalgame}
\bibliographystyle{apsrev4-1}
\end{document}